\input{aipcheck}

\edef\optionlist{%
   \variorefoptionifavailable        
   draft,%
   \psnfssproblemoption              
   tnotealph}

\documentclass[\optionlist]{aipproc}

\newcommand{\lbl}[1]{\label{eq:#1}}
\newcommand{ \rf}[1]{(\ref{eq:#1})}

\newcommand{\be}{\begin{equation}}
\newcommand{\ee}{\end{equation}}
\newcommand{\bea}{\begin{eqnarray}}
\newcommand{\eea}{\end{eqnarray}}
\newcommand{\setl}{\setlength\arraycolsep{2pt}}

\newcommand{\noi}{\noindent}

\newcommand{\ra}{\rightarrow}
\newcommand{\Ra}{\Rightarrow}

\newcommand{\lesssim}{ {\
\lower-1.2pt\vbox{\hbox{\rlap{$<$}\lower5pt\vbox{\hbox{$\sim$}}}}\ } 
}
\newcommand{\gtrsim}{ {\
\lower-1.2pt\vbox{\hbox{\rlap{$>$}\lower5pt\vbox{\hbox{$\sim$}}}}\ } 
}

\newcommand{\cL}{{\cal L}}

\newcommand{\tr}{\mbox{\rm tr}}

\newcommand{\MeV}{\mbox{\rm MeV}}
\newcommand{\GeV}{\mbox{\rm GeV}}

\newcommand{\with}{\mbox{\rm with}}

\newcommand{\annd}{\mbox{\rm and}}

\newcommand{\GF}{G_{\mbox{\rm {\tiny F}}}}

\newcommand{\eff}{\mbox{\rm {\footnotesize}eff}}

\newcommand{\stern}{\langle\bar{\psi}\psi\rangle}

\input epsf

\layoutstyle{6x9}

\usepackage{ttct0001}

\usepackage{shortvrb}
\MakeShortVerb\|

\makeatother

\begin{document}

\title
      [Large--$N_c$ QCD and Low Energy Interactions]
      {Large--$N_c$ QCD and Low Energy Interactions}


\author{Eduardo de Rafael}{
  address={\centerline{{\rm Centre de Physique ThŽorique
  }}CNRS--Luminy, case 907, F-13288 Marseille cedex 9, France },
  email={EdeR@cpt.univ-mars.fr},
  thanks={QCD&Work Conference BARI, Italy.}
}

\copyrightyear  {2001}

\begin{abstract}
This talk  reviews  recent progress in formulating the dynamics of
the electroweak interactions of hadrons at low energies, within the
framework of the $1/N_c$--expansion in QCD. The emphasis is put on 
the basic
issues of the approach.
\end{abstract}

\date{\today}

\maketitle

\section{Introduction}

In the Standard Model, the electroweak interactions of hadrons at
very low energies can be described by an effective Lagrangian which
only has as active degrees of freedom the flavour $SU(3)$ octet of
the low--lying pseudoscalar particles. The underlying theory,
however, is the gauge theory $SU(3)_{C}\times SU(2)_{L}\times
U(1)_{Y_w}$ which has as dynamical degrees of freedom quarks and
gauge fields. Going from these degrees
of freedom at high energies to an effective description in terms of
mesons at low energies is, in principle, a problem which
should be understood in terms of the evolution of the renormalization
group from short--distances to long--distances. Unfortunately, it is
difficult to carry out explicitly this evolution because at energies,
typically of a few
$\GeV$, non--perturbative dynamics like spontaneous chiral
symmetry breaking and color confinement sets in. 

It has been possible, however, to
integrate out the heavy degrees of
freedom of the Standard Model gauge theory, in the presence of the
strong interactions, perturbatively, thanks to the asymptotic
freedom property of QCD at short--distances. This procedure results in
an effective Lagrangian which consists of the usual QCD Lagrangian
with the light quark flavours $u$, $d$, and $s$ still active, plus a
sum of effective four--quark operators of the light quarks, modulated
by c--number coefficients (the Wilson coefficients,) which are
functions of the masses of the heavy particles which have been
integrated out, and of the renormalization scale. We are still left
with the evolution from this effective field theory, appropriate at
intermediate scales higher than a few
$\GeV$, to an effective Lagrangian description in terms of the low--lying
pseudoscalar particles which are the Goldstone modes associated to the
spontaneous chiral symmetry breaking of the Standard Gauge Theory in the
light quark sector. In this talk, I shall review recent  progress which has
been made in approaching this last step, when the problem is formulated
within the framework of QCD in the limit of a large number of colours
$N_c$. The emphasis is put on basic issues. Details of the applications
reviewed here can be found in the original publications.    

The suggestion to keep  $N_c$ as a free
parameter was first made by G.~'t Hooft~\cite{THFT74} as a possible
way to approach the study of non--perturbative aspects of QCD. The
limit $N_c\ra\infty$ is taken with the product $\alpha_{\mbox{\rm
{\footnotesize s}}} N_c$ kept fixed and it is highly non--trivial. In
spite of the efforts of many  illustrious theorists who have worked
on the subject, QCD in the large--$N_c$ limit still remains unsolved;
but many interesting properties have been proved, which suggest that,
indeed, the theory in this limit has the bulk of the
non--perturbative properties of the full QCD. In particular, it has
been shown that, if confinement persists in this limit, there is
spontaneous chiral symmetry breaking~\cite{CW80}. 

The spectrum of the
theory in the large--$N_c$ limit consists of an infinite number of
narrow stable meson states which are flavour nonets~\cite{W79}.
This spectrum looks {\it a priori} rather
different to the real world. The vector and axial--vector
spectral functions measured in $e^+ e^- \ra$ hadrons and in the
hadronic
$\tau$--decay show indeed a richer structure than just a sum of
narrow states. There are, however, many instances 
where one is only interested in observables which are given by weighted
integrals of some hadronic spectral functions. In these cases, it may
be enough to know a few {\it global} properties of the hadronic
spectrum, and to have a good interpolation. Typical examples of that
are, as we shall see, the coupling constants of the effective chiral
Lagrangian of QCD at low energies, as well as the coupling constants
of the effective chiral Lagrangian of the electroweak interactions of
pseudoscalar particles in the Standard Model. Some of these
couplings are needed in order to understand
$K$--Physics quantitatively. In these examples the
{\it hadronic world} predicted by large--$N_c$ QCD  provides already
a good approximation to the real hadronic spectrum. It is in this sense
that I shall show that large--$N_c$ QCD is a very useful
phenomenological approach for understanding non--perturbative QCD
physics at low energies.

There are a number of good articles  and lecture notes on large--$N_c$
QCD in the literature~\footnote{See e.g., the book in ref.~\cite{BW93}
and the lectures in \cite{Man99}}. Here I shall limit myself to make a
couple of comments on prejudices one often encounters concerning the
QCD large--$N_c$ limit.

\begin{itemize}

\item The first prejudice has to do with the ``extrapolation'' from
$N_c=3$ to $N_c=\infty$. In fact, $N_c$ is really used as a label
to select specific topologies among Feynman diagrams. The 
topology which corresponds to the highest power in the $N_c$--label is
the one which selects {\it planar} diagrams only; and the claim is
that it is this class that already provides a good approximation to
the full theory.

\item The second prejudice has to do with the fact that some physical
quantities are absent in the large--$N_c$ limit; the $\eta'$--mass
e.g. is zero in that limit. That does not mean that the
$1/N_c$--expansion fails, as sometimes it is argued, but rather that
some observables only appear at subleading topologies, ({\it planar}
diagrams with {\it one handle} in this case,) much the same as in
QED, there is no light--by--light scattering at the Born
approximation and one has to go to {\it one loop} diagrams to
evaluate its leading behaviour.
   
\end{itemize}

\section{The Chiral Lagrangian at Low Energies}

The strong and electroweak interactions of the Goldstone modes at
very low energies are described by an effective Lagrangian which has
terms with an increasing number of derivatives (and quark masses if
explicit chiral symmetry breaking is taken into account.) These terms
are modulated by couplings which encode the dynamics of the
underlying theory. The evaluation of these couplings from the
underlying theory is the question we are interested in. Typical terms
of the chiral Lagrangian are
$$
 \cL_{\eff}   =   \underbrace{\frac{1}4F_{0}^2\
\tr\left(D_{\mu}U
D^{\mu}U^{\dagger}\right)}_{ 
\pi\pi\ra\pi\pi\,,\quad K\ra\pi e\nu
}+\underbrace{L_{10}
\tr\left(
U^{\dagger}F_{R\mu\nu}
UF_{L}^{\mu\nu}\right)}_{ \pi\ra
e\nu\gamma }+\cdots 
$$
\be\lbl{chiral} 
\underbrace{e^{2}C
\tr\left(
Q_{R}UQ_{L}U^{\dagger}\right)}_{
-e^2 C\frac{2}{F_{0}^{2}}\left(\pi^{+}\pi^{-}+K^{+}K^{-}
\right)+\cdots}+\cdots -
\underbrace{\frac{\GF}{\sqrt{2}}V_{ud}V^{*}_{us} 
\ g_{8}F_{0}^{2}\left(D_{\mu}U
D^{\mu}U^{\dagger}\right)_{23}}_{
{K\ra\pi\pi\,, \quad K\ra\pi\pi\pi\,,\quad+\cdots}}+\cdots \,,
\ee
where $U$ is a $3\times 3$ unitary matrix in flavour space which
collects the Goldstone fields and which under chiral rotations
transforms as $U\ra V_{R}U V_{L}^{\dagger}$. The first line shows
typical terms of the strong interactions in the presence of external
currents~\cite{We79},\cite{GL84,GL85}; the second line shows typical
terms which appear when  photons, $W's$ and
$Z's$  have been integrated out in the presence of the strong
interactions. We show under the braces the typical physical processes
to which each term contributes. Each term is modulated by a constant:
$F_{0}^{2}$, $L_{10}$,... $C$...$g_{8}$... which encode the
underlying dynamics responsible for the appearance of the
corresponding effective term. Knowing $g_{8}$ for example, means that
we can calculate from first principles the dominant $\Delta I=1/2$
transitions for
$K$--decays to leading order in the chiral expansion. 

There are two
crucial observations concerning the relation of these low energy
constants to the underlying theory, that I
want to discuss.

\begin{itemize}

\item  The low--energy constants of the
Strong Lagrangian, like
$F_{0}^{2}$ and 
$L_{10}$,  are the
coefficients of the \underline{\it Taylor expansion}
of appropriate QCD Green's Functions. For example, with
$\Pi_{LR}(Q^2)$  the correlation function of a
left--current with a right--current in the chiral limit, (where the
light quark masses are neglected,) i.e.,
\be
2i\int d^4x\ e^{iq\cdot x}\langle 0\vert
T\left(\bar{u}_{L}\gamma^{\mu}d_{L}(x)
\bar{u}_{R}\gamma^{\nu}d_{R}(0)^{\dagger}\right)\vert
0\rangle =(q^{\mu}q^{\nu}-g^{\mu\nu}q^2)\Pi_{LR}(Q^2)\,,
\ee
the Taylor expansion  
\be 
-Q^2\Pi_{LR}(Q^2)
\vert_{Q^2\ra
0}=F_{0}^{2}-4L_{10}\ Q^2+
O\ (Q^4) \,,
\ee
defines the constants $F_{0}^{2}$ and $L_{10}$. 

\item By contrast, the low--energy constants 
of the ElectroWeak Lagrangian, like e.g.
$C$ and $g_{8}$, are  
\underline{\it integrals} of appropriate QCD Green's Functions. For
example, including the effect of weak neutral currents~\cite{KPdeR98},
\be\lbl{pimd}
C=\frac{3}{32\pi^2}\int_{0}^{\infty}dQ^2 \left(1-\frac{Q^2}{Q^2 +
M_{Z}^2}
\right)
\left(-Q^2\Pi_{LR}(Q^2)\right)\,.
\ee
Their evaluation appears to be,
a priori, quite a formidable task because they require the knowledge
of Green's functions at all values of the euclidean momenta; i.e.
they require a precise {\it matching} of the {\it short--distance}
and the {\it long--distance} contributions of the underlying Green's
functions.

\end{itemize}
 
\noi
The observations above are completely generic to the Standard
Model independently of the $1/N_c$--expansion. The
large--$N_c$ approximation helps, however, because it restricts the
{\it analytic structure} of the Green's functions in general, and
$\Pi_{LR}(Q^2)$ in particular, to be {\it meromorphic functions}: they
only have poles as singularities; e.g., in large--$N_c$ QCD,
\be\lbl{largeN}
\Pi_{LR}(Q^2)=\sum_{V}\frac{f_{V}^{2}M_{V}^2}{Q^2+M_{V}^2}-
\sum_{A}\frac{f_{A}^{2}M_{A}^2}{Q^2+M_{A}^2}-\frac{F_{0}^{2}}{Q^2}\,, 
\ee
where the sums are, in principle, extended to an infinite number of
states. 

There are two types of important restrictions on Green's
functions like $\Pi_{LR}(Q^2)$. One type follows from the fact
that, as already stated above, the Taylor expansion at low euclidean
momenta must match the low energy constants of the strong chiral
Lagrangian. This  results in a series of {\it\underline{long--distance
sum rules}} like e.g.
\be
\sum_{V}f_{V}^2- \sum_{A}f_{A}^2=-4L_{10}\,.
\ee

Another type of constraints follows
from the {\it \underline{short--distance  properties}} of the
underlying Green's functions. The behaviour at large euclidean momenta
of the Green's functions which govern the low energy constants of the
chiral Lagrangian in Eq.~\rf{chiral} can be obtained from the operator
product expansion (OPE) of local currents in QCD. In
the large--$N_c$ limit, this results in a series of algebraic sum
rules~\cite{KdeR98} which restrict the coupling constants and masses
of the hadronic poles. In the case of the $LR$--correlation
function in Eq.~\rf{largeN} one has,
{\setl 
\bea
{\mbox{\footnotesize\rm No $\frac{1}{Q^2}$ term in OPE $\Ra$ }} & 
\sum f_{V}^2M_{V}^2-\sum f_{A}^2M_{A}^2 -F_{0}^{2}=0\,,&
{\mbox{\footnotesize\rm 1st Weinberg sum rule. }} \\
{\mbox{\footnotesize\rm No $\frac{1}{Q^4}$ term in OPE $\Ra$ }} & 
\sum f_{V}^2M_{V}^4-\sum f_{A}^2M_{A}^4 =0\,, & 
{\mbox{\footnotesize\rm 2nd Weinberg sum rule. }} \\ \lbl{Q6}
{\mbox{\footnotesize\rm Matching$\frac{1}{Q^6}$ terms in the OPE $\Ra$
}} & \sum f_{V}^2M_{V}^6-\sum f_{A}^2 M_{A}^6
=\langle O^{(6)}\rangle\,, &
\eea}

\noi
where~\cite{KdeR98,SVZ79}, in large--$N_c$ QCD,
\be
\langle
O^{(6)}\rangle=\left[-4\pi\alpha_{s}+O(\alpha_{s}^2)\right]\stern^2\,.
\ee

\subsection{The Minimal Hadronic Ansatz Approximation to Large--$N_c$
QCD}

In most cases of interest, the Green's functions which govern the
low--energy constants of the chiral Lagrangian are order parameters of
spontaneous chiral symmetry breaking; i.e. they vanish, in the chiral
limit, order by order in the perturbative vacuum of QCD. That implies
that they have a power fall--out in $1/Q^2$ at large--$Q^2$; like e.g.,
the function
$\Pi_{LR}(Q^2)$, which as explicitly shown in Eq.~\rf{Q6}, falls as 
$1/Q^6$. That also implies that within a finite radius in the
complex $Q^2$--plane, these Green's functions in large--$N_c$ QCD, only have
a {\it finite number of poles}. The natural question which arises is:  {\sc
what is the minimal number of poles  required to satisfy the OPE
constraints?}  The answer to that follows from a well known theorem in
analysis~\cite{Tit39} which we illustrate with the example of the Green's
function
\be\lbl{delta}
-Q^2\Pi_{LR}(Q^2)\equiv\Delta[z]\quad\with\quad
z=\frac{Q^2}{M_{V_{1}}^2}\,,
\ee
where for convenience we normalize $Q^2$ to the mass of the lowest
vector state. The function $\Delta[z]$ has the property that
\be
N-P=\frac{1}{2\pi i}\oint\frac{\Delta'[z]}{\Delta[z]}dz\,,
\ee
where $N$ is the number of zeros and $P$ is the number of poles
inside the integration contour (a zero and/or a pole of order m is
counted m times.) For a contour of radius sufficiently large so that
the OPE applies, we simply have that
$N-P=-p_{\mbox{\rm{\footnotesize OPE}}}$
where $p_{\mbox{\rm{\footnotesize OPE}}}$ denotes the {\it
leading power} fall--out in $1/z$ predicted by the OPE. Since $N\ge 0$,
it follows that $P\ge p_{\mbox{\rm{\footnotesize OPE}}}$. In our
case~\footnote{Notice that with the definition of $\Delta[z]$ in
Eq.~\rf{delta} the pion pole is removed.}
$p_{\mbox{\rm{\footnotesize OPE}}}=2\Ra P\ge 2$  and the
{\it\underline {minimal hadronic ansatz}} (MHA) compatible with the
OPE requires two poles: one vector state and one axial--vector state.
The MHA approximation to the large--$N_c$ expression in
Eq.~\rf{largeN} is then the simple function
\be\lbl{MHA}
-Q^2\Pi_{LR}(Q^2)=F_{0}^{2}\frac{M_{V}^2
M_{A}^2}{(Q^2+M_{V}^2)(Q^2+M_{A}^2)}\,.
\ee
Inserting this function in Eq.~\rf{pimd} gives a prediction to the
low--energy constant $C$ which governs the electromagnetic
$\pi^{+}-\pi^{0}\equiv\Delta m_{\pi}$ mass difference, with the
result~\footnote{This is the result for $M_{V}=(748\pm 29)\,\MeV$ and
$g_{A}=\frac{M_{V}^2}{M_{A}^2}=0.50\pm 0.06$. These values follow from
an overall fit to predictions of the low energy
constants~\cite{PPdeR98,GP00}.}
\be\lbl{mpimha}
\Delta m_{\pi}=(4.9\pm 0.4)\,\MeV\,,\qquad {\mbox{\rm{
MHA to Large--$N_c$ QCD}}}\,, 
\ee
to be compared with the experimental value
\be
\Delta m_{\pi}=(4.5936\pm 0.0005)\,\MeV\,,\qquad {\mbox{\rm
Particle Data Book~\cite{PDG00}}}\,. 
\ee

The shape of the  function in Eq.~\rf{MHA}, normalized to its value at
$Q^2=0$, is shown in Fig.~1 below, (the continuous red curve.) It provides a
good interpolation between the low--$Q^2$ regime where
$\chi$PT applies and the high--$Q^2$ regime where the OPE applies.
Also shown in the same plot is the experimental curve, (the green
dotted curve) obtained from the ALEPH collaboration
data~\cite{ALEPH}. Except for the intermediate energy region, where
the MHA overestimates slightly the experimental curve, the overall
agreement is quite remarkable.

\centerline{\epsfbox{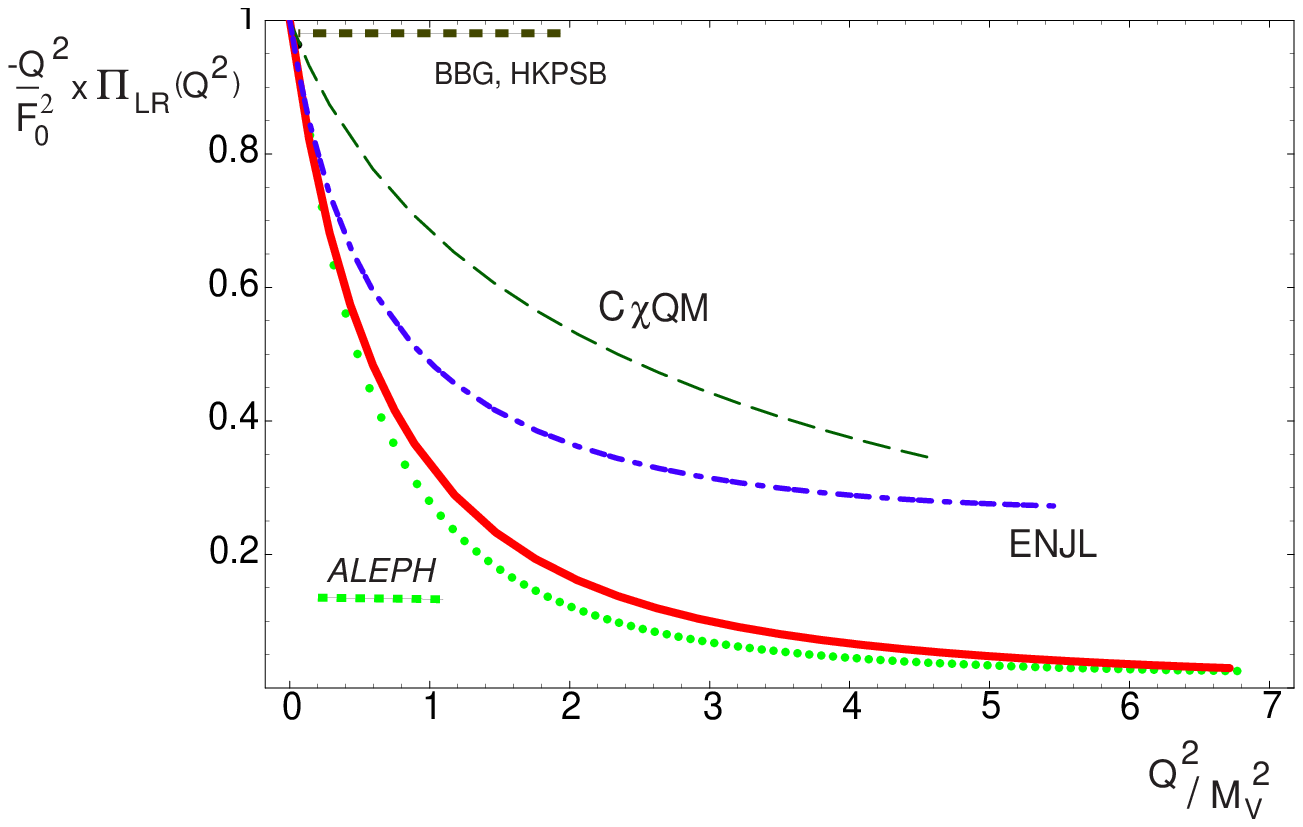}}
\begin{center}
{{\bf Fig.~1} {\it MHA 
(red curve,) versus  ALEPH (green dots) and other predictions}}
\end{center}

The MHA approximation to large--$N_c$ QCD is a starting point to do 
well defined approximations in nonperturbative QCD. The approximation
has been tested with the ALEPH data~\cite{PPdeR01}. In principle it is
improvable: inserting more terms in the OPE provides extra sum rules
which can be used to fix the extra hadronic parameters. We have made
tests with models of large--$N_c$ QCD~\cite{PPdeR01b,GP01}. It has
also been shown~\cite{Bthesis} that in the case of $\Pi_{LR}$,
inserting an extra $\rho'$--like vector state, improves the overall
picture; the
$\Delta m_{\pi}$ prediction in particular.

At this stage, it is perhaps illustrative to compare the large--$N_c$
approach we have discussed so far with other analytic
approaches which exist in the literature. The $\Pi_{LR}$
correlation function provides us with an excellent theoretical laboratory
to do that. The different shapes of this correlation function predicted by
other analytic approaches are also collected in Fig.~1. Let us comment on
them individually.

\begin{itemize}
\item The suggestion to use large--$N_c$ QCD combined with lowest order
$\chi$PT loops, was first proposed by Bardeen, Buras and G\'erard in a
series of seminal papers~\footnote{See refs.~\cite{Bu88,Ba89,Ge90}
and references therein.}. The same approach has been applied by the
Dortmund group~\cite{Dortmmund}, in particular to the evaluation of
$\epsilon'/\epsilon$. In this approach the {\it hadronic ansatz} to
the Green's functions consists of Goldstone poles only and their
integrals, (which become of course UV--divergent, often quadratically
divergent since the correct QCD short--distance behaviour is not
implemented,) are cut--off sharply. In this case, the predicted
hadronic shape of the LR--correlation function normalized to its
value at $Q^2=0$ is constant,
as shown by the BBG, HKPSB line (black dotted) in Fig.~1.

\item
The Trieste group  evaluate the relevant Green's functions using
the constituent chiral quark model (C$\chi$QM) proposed in
refs.~\cite{MG84} and \cite{EdeRT90,PdeR91}. They have obtained a long list
of  predictions~\cite{Trieste}, in
particular
$\epsilon'/\epsilon$. The model gives an educated first guess of the
low--$Q^2$ behaviour of the Green's functions, as one can judge from
the C$\chi$QM--curve (green dashed) in Fig.~1, but it 
fails to reproduce the short--distance QCD--behaviour. Another
objection to this approach is that the ``natural matching scale'' to
the short--distance behaviour in this model should be
$\sim 4M_{Q}^2$, ($M_{Q}$ the constituent quark mass,) too low to be
trusted.

\item
The extended Nambu--Jona-Lasinio (ENJL) model was developed as an
improvement on the C$\chi$QM, since in a certain way it incorporates
the vector--like fluctuations of the underlying QCD theory, which are
known to be phenomenologically important~\footnote{For a review, see
e.g. ref.~\cite{Bij96} where earlier references can be found.}.  The
model is, indeed, rather successful in predicting the low--energy
$O(p^4)$ constants of the chiral Lagrangian. It has, indeed, a better
low--energy behaviour than the C$\chi$QM, as the ENJL--curve
(blue dot--dashed) in Fig.~1 shows; but it  fails to
reproduce the short--distance behaviour of the OPE in QCD. Arguments to do
the matching to short--distance QCD have been forcefully  elaborated in
refs.~\cite{LuGr}, which also have made a lot of predictions; a
large value for
$\epsilon'/\epsilon$ in particular.

\item
The problem with the ENJL
model as a plausible model of large--$N_c$ QCD, is that the on--shell
production of unconfined constituent quark $Q\bar{Q}$ pairs that it
predicts violates the  large--$N_c$ QCD counting rules.
In fact, as shown in ref.~\cite{PPdeR98}, when the unconfining pieces
in the ENJL spectral functions are removed by adding an appropriate
series of local counterterms, the resulting theory is entirely
equivalent to an effective chiral meson theory with three narrow
states V, A and S; very similar to the phenomenological {\it Resonance
Dominance} Lagrangians proposed in refs.~\cite{EGPdeR89,EGLPdeR89}.
These {\it Resonance
Dominance} Lagrangians can be viewed as particular models of
large--$N_c$ QCD. They predict the same  Green's functions as the MHA
approximation to large--$N_c$ QCD discussed above, in 
\underline{some} particular cases but not in general~\footnote {See
e.g. the three--point functions discussed in ref.~\cite{KN01}.}.  
  
\end{itemize}

\noi
In view of the difficulties that these analytic approaches
have in reproducing the shape of  the simplest Green's
function one can think of, it is difficult to attribute more than a 
qualitative significance to their ``predictions'';
$\epsilon'/\epsilon$ in particular, which requires the interplay of
several other Green's functions much more complex than
$\Pi_{LR}(Q^2)$.

\section{ Methodology and Applications}

The approach that we propose in order to compute a specific coupling
of the chiral electroweak Lagrangian consists of the following steps:

\begin{enumerate}
\item
{\it Identify the underlying QCD Green's functions.}

In
most cases of interest, the Green's functions in question are
two--point functions with zero momentum insertions of vector,
axial vector, scalar and pseudoscalar currents. The higher the
power in the chiral expansion, the higher will be the number of
insertions. This step is totally general and does not invoke any
large--$N_c$ approximation. 

\item{\it Work out the short--distance behaviour and the
long--distance behaviour of the relevant Green's functions.}

The long--distance behaviour is governed by the Goldstone
singularities and can be obtained from $\chi$PT. The
short--distance behaviour is governed by the OPE of the currents
through which the hard momenta flows. Again, this step is well
defined independently of the large--$N_c$ expansion; in practice,
however, the calculations simplify a lot when restricted to the
appropriate order in the $1/N_c$--expansion one is interested in.

\item
{\it Large--$N_c$ ansatz for the underlying Green's functions.}

As already mentioned, the large--$N_c$ ansatz involves only sums of
poles; the {\it minimal hadronic ansatz} consists in limiting these
sums to the minimum number required to satisfy the 
leading power fall--out
at short--distances, as well as the appropriate $\chi$PT long--distance
constraints.  
\end{enumerate}

\noi
All the three steps can be done analytically which helps to show the
crucial points of the underlying dynamics. The method is, in principle,
\underline{improvable}~\footnote{Unlike the other analytic methods
discussed above.} by adding constraints from the next--to--leading
short--distance inverse power behaviour and/or higher orders in the chiral
expansion.  

We have tested this approach with the calculation of a few
low--energy observables:

\begin{itemize}
\item {\em The electroweak $\Delta m_{\pi}$ mass
difference}~\cite{KPdeR98} which we have already discussed. 

\item {\em The hadronic vacuum polarization contribution to the
anomalous magnetic moment of the muon
$a_{\mu}$}~\cite{PdeR}.
The MHA in this case requires one
vector--state pole and a pQCD continuum. The absence of dimension two
operators in QCD in the chiral limit, constrains the threshold of the
continuum. The result, which includes an estimate of the systematic
error of the approach, is 
\be
a_{\mu}\vert_{HVP}=(5.7\pm 1.7)\times 10^{-8}\,,
\ee
to be compared with an average of recent
phenomenological determinations~\footnote{See e.g. Prades's talk at
KAON2001~\cite{Pr01} and references therein.}
\be
a_{\mu}\vert_{HVP}=(6.949\pm 0.064)\times 10^{-8}.
\ee

\item
{\em The $\pi^{0}\ra e^{+}e^{-}$ and $\eta\ra \mu^{+}\mu^{-}$ decay
rates}~\cite{KPPdeR99}. These processes are governed by a $\langle
PVV\rangle$ three--point function, with the $Q^2$--momentum flowing through
the two $V$ insertions. The MHA in this case requires a vector--pole and a 
double vector--pole. The predictions of the branching ratios
\be\lbl{rat}
R(P\ra l^{+}l^{-})=\frac{\Gamma(P\ra l^{+}l^{-})}{\Gamma(P\ra
\gamma\gamma)}\,,
\ee
compared to the experimental determinations are shown in Table~1 below.
\begin{table}[h]
\caption{\em Summary of branching ratios results}
\begin{tabular}{@{}l@{}l@{}l@{}}
\hline\hline
{\bf Branching Ratio}~\rf{rat}~~~ & {\bf Large--$N_c$ Approach}~~~~~~~~~~ &
{\bf Experiment}~\cite{PDG00}\\
\hline
$R(\pi^0\to e^+e^-)\times 10^{8}\qquad$ & $6.2\pm 0.3\qquad$ & 
$6.28\pm 0.55$
  \\ \hline
$R(\eta\to \mu^+\mu^-)\times 10^{5}$ & $1.4\pm 0.2$ & $1.47\pm 0.20
$ \\ \hline
$R(\eta\to e^+e^-)\times 10^{8}$ & $1.15\pm 0.05$ &
 $< 1.8\times 10^{4}$
\\
\hline\hline
\end{tabular}
\end{table}  

\end{itemize}

\noi
These successful predictions have encouraged us to start a
project of a systematic analysis of non--leptonic $K$--decays within
this large--$N_c$ approach. We have first used the example of the
neutral current contribution to the $\Delta m_{\pi}$ mass
difference, (see Eq.~\rf{pimd},) as a theoretical laboratory to show
explicitly the cancellation between the
renormalization scale in the Wilson coefficient of four--quark
operators and in the hadronic matrix elements evaluated in our
approach. We have later shown that this cancellation can also be
made renormalization scheme independent~\cite{PdeR00,KPdeR01}. 

So far we have completed two calculations of $K$--matrix elements
within this large--$N_c$ approach, which we next discuss. 

\subsection{The $B_{K}$--Factor of $K^{0}-\bar{K}^{0}$ Mixing}

The factor in question is conventionally defined by the matrix
element of the four--quark operator $Q_{\Delta S=2}(x)=
(\bar{s}_{L}\gamma^{\mu}d_{L})
(\bar{s}_{L}\gamma_{\mu}d_{L})(x)$:
\be
\langle \bar{K}^{0}\vert Q_{\Delta S=2}
(0)\vert
K^{0}\rangle =\frac{4}{3}f_{K}^2 M_{K}^2 B_{K}(\mu)\,.
\ee
To lowest order in the chiral expansion the operator  $Q_{\Delta
S=2}(x)$ bosonizes into a term of $O(p^2)$
\be
Q_{\Delta S=2}(x)\Ra -\frac{F_{0}^4}{4}g_{\Delta S=2}(\mu)
\left[(D^{\mu}U^{\dagger})U\right]_{23}
\left[(D_{\mu}U^{\dagger})U\right]_{23}\,,
\ee
with $g_{\Delta S=2}(\mu)$ a low energy constant, to be determined, 
which is a function of the renormalization scale $\mu$ of the
Wilson coefficient $C(\mu)$ which modulates the operator $Q_{\Delta
S=2}(x)$ in the four--quark effective Lagrangian. A convenient
choice of the underlying Green's function here is the four--point
function
$W_{LRLR}(Q^2)$ of two left--currents which carry the
$Q^2$--momentum one has to integrate over, and two right--currents
with zero momentum insertions. The coupling constant $g_{\Delta
S=2}(\mu)$, which has to be evaluated in the same renormalization
scheme as the Wilson coefficient $C(\mu)$ has been calculated, is then
given by an integral~\cite{PdeR00}
\be\lbl{bkint}
g_{\Delta S=2}(\mu)=1-\frac{1}{32\pi^2 F_{0}^2}\int_{0}^{\infty}
dQ^2\left(\frac{4\pi\mu^2}{Q^2} \right)^{\epsilon/2}W_{LRLR}(Q^2)\,,
\ee 
conceptually similar to the one which determines the electroweak
constant
$C$ in Eq.~\rf{pimd}. The {\it hadronic ansatz}, in the
$1/N_c$--expansion, of the Green's
function
$W_{LRLR}(Q^2)$ ,  which fulfills the leading short--distance
constraint and the long--distance constraints which fix  $W_{LRLR}(0)$
and
$W_{LRLR}'(0)$, requires one vector--pole, a double vector--pole and
a triple vector--pole~\footnote{This goes beyond the strict MHA
which, in this case,  only requires a vector--pole. It is the
{\it extra} information of knowing $W_{LRLR}(0)$
and
$W_{LRLR}'(0)$ in $\chi$PT which forces the presence of the double and
triple poles.}. The invariant
$\hat{B}_{K}$  defined as
\be
\hat{B}_{K}=\frac{3}{4}C(\mu)\times g_{\Delta S=2}(\mu)\,,
\ee
can then be evaluated, with no free parameters, with the
result~\cite{PdeR00}
\be\lbl{bk}
\hat{B}_{K}=0.38\pm 0.11\,.
\ee
When comparing this result to other determinations, specially in
lattice QCD, it should be realized that the unfactorized 
contribution in Eq.~\rf{bkint} is the one in the chiral limit.
It is possible, in principle, to calculate chiral corrections within
the same large--$N_c$ approach, but this has not yet been done. 

The result in Eq.~\rf{bk} is compatible with 
the old current algebra prediction~\cite{DGH82} which, 
to lowest order in
chiral perturbation theory, relates the
$B_{K}$ factor to the $K^{+}\ra \pi^{+}\pi^{0}$ decay rate. 
In fact, our
calculation of the $B_{K}$ factor can be viewed as a 
successful prediction of
the $K^{+}\ra \pi^{+}\pi^{0}$ decay rate!  

As discussed in ref.~\cite{PdeR96}
the bosonization of the four--quark operator
$Q_{\Delta S=2}$ and the bosonization of the operator
$Q_{2}-Q_{1}$ which generates $\Delta I=1/2$ transitions  
are related to each
other in the combined chiral limit and  next--to--leading 
order in the $1/N_c$--expansion. Lowering the value of $\hat{B}_{K}$ from
the leading large--$N_c$ prediction $\hat{B}_{K}=3/4$ to the result in 
Eq.~\rf{bk} is
correlated with an increase of the coupling constant $g_{8}$ 
in the lowest order
effective chiral Lagrangian, (see Eq.~\rf{chiral},) which generates 
$\Delta I=1/2$ transitions, and
provides a first step towards a quantitative understanding of 
the dynamical
origin of the  $\Delta I=1/2$ rule.

\subsection{ ElectroWeak Four--Quark Operators}

These are the four--quark operators generated by the so called electroweak
Penguin like diagrams~\footnote{See e.g., Buras lectures\cite{Bu99}}
\be
\cL\Ra \cdots
C_{7}(\mu)Q_{7}+
C_{8}(\mu)
Q_{8}\,,
\ee
with
\be
Q_7 = 6(\bar{s}_{L}\gamma^{\mu}d_{L})
\sum_{q=u,d,s} e_{q} (\bar{q}_{R}\gamma_{\mu}q_{R})
\quad\annd\quad Q_8 =
-12\sum_{q=u,d,s}e_{q}(\bar{s}_{L}q_{R})(\bar{q}_{R}d_{L})\,, 
\ee
and $C_{7}(\mu)$, $C_{8}(\mu)$ their corresponding Wilson coefficients.
They generate a term of $O(p^0)$ in the effective chiral
Lagrangian\cite{BW84}; therefore, the matrix elements of
these operators, although suppressed by an $e^2$ factor, are chirally
enhanced. Furthermore, the Wilson coefficient $C_{8}$ has a large
imaginary part, which makes the matrix elements of the
$Q_{8}$ operator to be particularly important in the evaluation of
$\epsilon'/\epsilon$. 

Within the large--$N_c$ framework, the bosonization of these
operators produce matrix elements with the following counting
\be\lbl{Q7Q8z}
\langle Q_{7}\rangle\vert_{O(p^0)}= \underline{O(N_c)}      
+O(N_c^0)
\quad\annd\quad
\langle Q_{8}\rangle\vert_{O(p^0)}= \underline{O(N_{c}^2)}
+\!\!\!\!\!\!\!\!\!
\underbrace{O(N_c^0)}_{{\mbox{\rm
Zweig suppressed}}} 
\ee
A first estimate of the underlined contributions was made in
ref.~\cite{KPdeR99}~\footnote{The inclusion of final state interaction
effects based on the leading large--$N_c$ determination of $\langle
Q_{8}\rangle$, (and $\langle
Q_{6}\rangle$,) in connection with a phenomenological determination of
$\epsilon'/\epsilon$, has been recently discussed in ref.~\cite{PP01}.}. The
bosonization of the
$Q_{7}$ operator to
$O(p^0)$ in the chiral expansion and to $O(N_c)$ is very similar to
the calculation of the $Z$--contribution to the coupling constant $C$
in Eq.~\rf{pimd}. An evaluation which also takes into account the
renormalization scheme dependence has been recently made in
ref.~\cite{KPdeR01}. 

The contribution of $O(N_c^0)$ to 
$\langle Q_{8}\rangle\vert_{O(p^0)}$ is
Zweig suppressed. It involves the sector of 
scalar (pseudoscalar)
Green's functions  where it is hinted from various 
phenomenological sources that
the restriction to just the 
leading large--$N_c$ contribution may not always be a good
approximation. Fortunately, as first pointed out by the authors of ref.
~\cite{DG00}, independently of large--$N_c$ considerations, the
bosonization of the
$Q_8$ operator to $O(p^0)$ in the chiral expansion 
can be related to the four--quark condensate 
$\langle O_2 \rangle\equiv
\langle 0\vert (\bar{s}_{L}s_{R})(\bar{d}_{R}d_{L})\vert
0\rangle$ by current
algebra Ward identities, the same four--quark condensate which also
appears in the OPE of the $\Pi_{LR}(Q^2)$ function  discussed above.
More precisely
\be\lbl{ope2}
\lim_{Q^2\ra\infty} \left(
-Q^2\Pi_{LR}(Q^2)\right)Q^4  = 
4\pi^2\frac{\alpha_{s}}{\pi}\left(4\langle
O_{2}\rangle+\frac{2}{N_c}\langle O_{1}\rangle\right)+O
\left(\frac{\alpha_{s}}{\pi}\right)^2\,,
\ee 
where $\langle O_1\rangle \equiv\langle 0\vert
(\bar{s}_{L}\gamma^{\mu}d_{L})(\bar{d}_{R}\gamma_{\mu}s_{R})\vert 0 
\rangle $ is the vev which governs $\langle
Q_{7}\rangle\vert_{O(p^0)}$. In fact, in the
$1/N_c$--expansion~\cite{KPdeR01}
{\setl
\bea\lbl{LRgralmsb}
\langle O_1\rangle & = & \left(-\frac{1}{2} ig_{\mu\nu}\int
\frac{d^4q}{(2\pi)^4}
\Pi_{LR}^{\mu\nu}(q)\right)_{\overline{\mbox{\rm{\footnotesize
MS}}}}^{\mbox{\rm\footnotesize ren.}}  \\&  = &  
 -\frac{3}{32\pi^2}\left[\sum_{A}f_{A}^2
M_{A}^{6}\log\frac{\Lambda^2}{M_{A}^2}-\sum_{V}f_{V}^2
M_{V}^{6}\log\frac{\Lambda^2}{M_{V}^2}
\right]\,,
\eea}

\noi
with (NDR means naive dimensional renormalization scheme; HV means 't
Hooft--Veltman scheme,) 
\be\lbl{cutoff}
\Lambda^2=\mu^2\exp(1/3+\kappa)\,;\qquad
\kappa=-1/2\quad\mbox{\rm in NDR}\,,\quad\annd\quad
\kappa=+3/2\quad\mbox{\rm
in HV}\,.
\ee
The crucial observation is that large--$N_c$ QCD gives a rather good
description of the $\Pi_{LR}(Q^2)$--function, as we have seen earlier; in
particular it implies that, (see Eq.~\rf{Q6},)
\be
\lim_{Q^2\ra\infty} \left(
-Q^2\Pi_{LR}(Q^2)\right)Q^4=\sum_V f_V^2 M_V^6 - \sum_A f_A^2 M_A^6 \,.
\ee
Solving these equations in the MHA approximation, results in a
determination of the matrix elements of
$\langle Q_{8}\rangle\vert_{O(p^0)}$ which does not require the
separate knowledge of the Zweig suppressed $O(N_c^{0})$ term in
Eq.~\rf{Q7Q8z}. 

The numerical results we get for the matrix elements
\be\lbl{M78}
M_{7,8}\equiv \langle (\pi\pi)_{I=2}\vert Q_{7,8}\vert
K^{0}\rangle_{~2\,
\GeV}
\ee
at the renormalization scale $\mu=2~\GeV $ in the two schemes NDR
and HV and in units of
$\GeV^{3}$ are shown in Table~2 below, (the first line.)
\vskip 1pc
\begin{table}[h]
\caption{\em Matrix elements  results, (see Eq.~\rf{M78})}
\begin{tabular}{@{}l@{}l@{}l@{}l@{}l@{}}
\hline\hline
METHOD & $M_{7}$(NDR)~~~~~~ & $M_{7}$(HV)~~~~~~~ &
$M_{8}$(NDR)~~~~~~~ &
$M_{8}$(HV)~~~~~~\\
\hline
{\bf Large--$N_c$ Approach} & & & & \\
Ref.~\cite{KPdeR01} & $0.11\pm 0.03$ & $0.67\pm 0.20$ & $3.5\pm 1.1$
& $3.5\pm 1.1$ \\
\hline
{\bf Lattice QCD} & & & & \\
Ref.~\cite{Donetal} & $0.11\pm 0.04$ & $0.18\pm 0.06$ & $0.51\pm 0.10$
& $0.62\pm 0.12$ \\
\hline
{\bf Dispersive Approach} & & & & \\
Ref.~\cite{DG00}~~~~~~ & $0.22\pm 0.05$ &  & $1.3\pm 0.3$
& \\
Ref.~\cite{Na01} & $0.35\pm 0.10$ &  & $2.7\pm 0.6$ &  \\
Ref.~\cite{Go01}~~~~~~~~~& $0.18\pm 0.12$~~~~~&
$ 0.50\pm 0.06 $~~~~~& 
$2.13\pm 0.85$~~~~~~~ & $2.44\pm 0.86$ \\
Ref.~\cite{CDGM01} & $0.16\pm 0.10$ & $0.49\pm 0.07$ & $2.22\pm 0.67$ &
$2.46\pm 0.70$  \\
\hline\hline
\end{tabular}
\end{table}  
 
\noi
Also shown in the same table
are other 
evaluations of matrix elements with which we can compare scheme 
dependences
explicitly~\footnote{There is another "dispersive determination" in the
literature~\cite{BGP01} since the HEP-2001 conference, but it is
controversial as yet; this is why we do not include it in the Table.}.
Several remarks are in order

\begin{itemize}

\item  Our evaluations of $M_8$ do not include the terms of
$O(\alpha_{s}^2)$ in Eq.~\rf{ope2} because, as pointed out in
Ref.~\cite{KPdeR01}, the available results in the literature~\cite{LSC86}
were not calculated in the right basis of four--quark operators needed
here.  

\item We find
that our results for $M_{7}$ are in very good
agreement with the lattice results in the NDR scheme, but not in the HV
scheme. This disagreement is, very likely, correlated with the strong
discrepancy we have with the lattice result for $M_{8}$(NDR).

\item
The recent revised dispersive approach results~\cite{Go01,CDGM01}, which now
include the  effect of higher
terms in the  OPE, are in agreement, within errors, with
the large--$N_c$ approach results. In fact, the agreement improves further
if the new
$O(\alpha_{s}^2)$ corrections, which have now been calculated in the right
basis~\cite{CDGM01}, are also incorporated in the large--$N_c$ approach.

\item
Both the   revised dispersive approach results and the 
large--$N_c$ approach
results for $M_{8}$ are higher than the lattice results. The 
discrepancy may originate in the fact that, for reasonable 
values of $\stern$, 
most of the contribution to  $M_{8}$ appears to come from an OZI--violating
Green's function which is something inaccessible in the quenched
approximation at which the lattice results, so far, have been obtained. 
\end{itemize}

\subsection{Conclusion and Outlook}

We hope to have shown with these examples that the 
large--$N_c$ approach that we
have discussed, provides a very useful framework to formulate 
calculations of
genuinely non--perturbative nature, like the low--energy constants of the
effective chiral Lagrangian, both in the strong and the electroweak sector.

Other calculations in progress, by various groups of
people and  in order of advancement, are
\begin{itemize}
\item {\em The electroweak hadronic contributions to  $g_{\mu}-2$.}

\item {\em The matrix elements of the strong Penguin operator $Q_6$}, 
relevant for
$\epsilon'/\epsilon$.

\item {\em The light--by--light hadronic contributions to $g_{\mu}-2$}; in
particular the one  generated by the convolution of two $\langle PVV\rangle$
three--point functions.

\item {\em The chiral corrections to the $B_{K}$--factor.}
\end{itemize}

\noi
We hope to have the results in the near future.

\section{Acknowledgements}

My knowledge on the subject reported here owes much to enjoyable discussions and
work with Marc Knecht, Santi Peris, Michel
Perrottet, and Toni Pich, as well as, more recently,  with
Thomas Hambye, Laurent Lellouch, Andreas Nyffeler and Boris Phily. I wish
to thank them all here.


\begin{thebibliography}{99}

\bibitem{THFT74}
         G.~'t Hooft, Nucl. Phys. {\bf B72} (1974) 461; {\bf B73}
         (1974) 461.

\bibitem{CW80}
         S.~Coleman and E.~Witten, Phys. Rev. Lett., {\bf 45} (1980)
         100.

\bibitem{W79}
         E.~Witten, Nucl. Phys. {\bf B160} (1979) 57. 

\bibitem{BW93}
         {\it The large N Expansion in Quantum Field Theory and
          Statistical Physics}, Editors E.~Brezin and S.R.~Wadia,
          World Scientific, 1993.

\bibitem{Man99}
         A.V.~Manohar, {\it Large--$N_c$ QCD}, in les Houches Session
         LXVIII, North Holland,
         1999.

\bibitem{We79}
         S.~Weinberg, Physica {\bf A96} (1979) 327.

\bibitem{GL84}
         J.~Gasser and H.~Leutwyler, Ann. of Phys.(N.Y.) {\bf 158}
         (1984) 142.

\bibitem{GL85}
         J.~Gasser and H.~Leutwyler, Nucl. Phys.  {\bf B250} (1985)
         465.

\bibitem{KPdeR98}
         M.~Knecht, S.~Peris and E.~de Rafael, Phys. Lett. {\bf B443}
         (1998) 255.

\bibitem{KdeR98}
         M.~Knecht and E.~de Rafael, Phys. Lett. {\bf B424} (1998)
         355.

\bibitem{SVZ79}
         M.A.~Shifman, A.I.~Vainshtein and V.I.~Zakharov, Nucl. Phys.
         {\bf B147} (1979) 385, {\it ibid} 447.

\bibitem{Tit39}
         E.C.~Titchmarsh, {\it The Theory of Functions}, 2nd edition,
         OUP 1939.

\bibitem{PPdeR98}
         S.~Peris, M.~Perrottet and E.~de Rafael, JHEP {\bf 05} 
									(1998) 011.

\bibitem{GP00}   
         M.~Golterman and S.~Peris, Phys. Rev. {\bf D61}
         (2000) 034018.

\bibitem{PDG00}
									{\it Review of Particle Physics}, Eur. Phys. J. {\bf C15}
         (2000) 1.

\bibitem{ALEPH}
         ALEPH Collaboration, R.~Barate {\it et al}, Z. Phys. {\bf C76}
         (1997) 15; {\it ibid} Eur. Phys. J. {\bf C4} (1998) 409.
 
\bibitem{PPdeR01}
         S.~Peris, B.~Phily and E.~de Rafael, Phys. Rev. Lett. {\bf
         86} (2001) 14.

\bibitem{PPdeR01b}
         S.~Peris, B.~Phily and E.~de Rafael, {\it talk at the
         EURODAPHNE--Marseille meeting Feb.~2001} to be published.

\bibitem{GP01}
         M.~Golterman and S.~Peris, JHEP {\bf 01} (2001) 028.

\bibitem{Bthesis}
         B.~Phily, PhD--Thesis, University of Marseille, Luminy.

\bibitem{Bu88}
         A.~Buras, {\it The $1/N$ approach to nonleptonic weak
interactions}, in
          CP violation, ed. C.~Jarlskog, World Scientific, Singapore,
          1998.

\bibitem{Ba89}
         W.~Bardeen, {\it Weak decay amplitudes in large $N_c$ QCD},
         in Proc. of Ringberg Workshop, Nucl. Phys.B (Proc. Suppl.)
         {\bf 7A} (1989) 149. 
        
\bibitem{Ge90}
         J.-M.~G\'erard, Acta Phys. Pol. {\bf B21} (1990) 257.

\bibitem{Dortmmund}
         T.~Hambye, G.O.~K\"ohler, E.A.~Paschos, P.H.~Soldan and
         W.A.~Bardeen,
         Phys. Rev. {\bf D58} (1998) 014017, and references therein.

\bibitem{MG84}
         A.~Manohar and H.~Georgi, Nucl. Phys. {\bf B234} (1984) 189.

\bibitem{EdeRT90}
         D.~Espriu, E.~de Rafael and J.~Taron, Nucl. Phys. {\bf
         B345} (1990) 22.

\bibitem{PdeR91}
         A.~Pich and E.~de Rafael, Nucl. Phys. {\bf B358} (1991) 311.

\bibitem{Trieste}
         S.~Bertolini, M.~Fabbrichesi and J.O.~Egg, Rev. Mod. Phys. {\bf 72}
         (2000) 65 and references therein.  

\bibitem{Bij96}
         J.~Bijnens, Phys. Rep. {\bf 265(6)} (1996) 369. 

\bibitem{LuGr}
         J.~Bijnens and J.~Prades, JHEP {\bf 9901} (1999) 023 ; 
									{\it ibid} {\bf
         0001} (2000) 002; {\it ibid} {\bf 0006} (2000) 035.

\bibitem{EGPdeR89}
         G.~Ecker, J.~Gasser, A.~Pich and E.~de Rafael, Nucl. Phys.
         {\bf B321} (1989) 311.

\bibitem{EGLPdeR89}
          G.~Ecker, J.~Gasser, H.~Leutwyler, A.~Pich and E.~de Rafael,
          Phys. Lett. {\bf B321} (1989) 425.

\bibitem{KN01}
          M.~Knecht and A.~Nyffeler, hep-ph/0106034.

\bibitem{PdeR}
         M.~Perrottet and E.~de Rafael, {\it unpublished}.

\bibitem{Pr01}
         J.~Prades, hep-ph/0108192.

\bibitem{KPPdeR99}
         M.~Knecht, S.~Peris, M.~Perrottet and E.~de Rafael, Phys.
         Rev. Lett. {\bf 83} (1999) 5230.

\bibitem{PdeR00}
         S.~Peris and E.~de Rafael, Phys. Lett. {\bf B490} (2000) 213,
         {\it erratum} hep-ph/0006146 v3.

\bibitem{KPdeR01}
         M.~Knecht, S.~Peris and E.~de Rafael, Phys. Lett. {\bf B508}
         (2001) 117. 

\bibitem{DGH82}
         J.F.~Donoghue, E.~Golowich and B.R.~Holstein, Phys. Lett. 
         {\bf B119} (1982) 412.

\bibitem{PdeR96}
         A.~Pich and E.~de Rafael, Phys. Lett. {\bf B374} (1996) 186.

\bibitem{Bu99}
         A.J.~Buras, {\it Weak Hamiltonian, CP Violation and Rare
         Decays}, in les Houches Session
         LXVIII, North Holland,
         1999.

\bibitem{BW84}
         J.~Bijnens and M.~Wise, Phys. Lett. {\bf B137} (1984) 245.

\bibitem{KPdeR99}
         M.~Knecht, S.~Peris and E.~de Rafael, Phys. Lett. {\bf B457}
         (1999) 227.

\bibitem{PP01}
         E.~Pallante, A.~Pich and I.~Scimemi, hep-ph/0105011 and references
         therein.

\bibitem{LSC86}
         L.V.~Lanin, V.P.~Spiridonov and K.~G.~Chetyrkin, Sov. J. Nucl.
         Phys. {\bf 44} (1986) 896.

\bibitem{Donetal}
         A.~Donini {\it et al}, Phys. Lett. {\bf B470} (1999) 233.

\bibitem{DG00}
         J.F.~Donoghue and E.~Golowich, Phys. Lett. {\bf B478} 
         (2000) 172. 

\bibitem{Na01}
         S.~Narison, Nucl. Phys.  {\bf B593} (2001) 3;
         hep-ph/0012019;   {\it and private communication.} 

\bibitem{Go01}
         E.~Golowich, Talk at the EPS-HEP-2001 Conference, Budapest.

\bibitem{BGP01}
									J.~Bijnens, E.~G\'amiz and J.~Prades, hep-ph/0108240, submitted to
         JHEP. 

\bibitem{CDGM01}
         V.~Cirigliano, J.~Donoghue, E.~Golowich and K.~Maltman,
         hep-ph/0109113 v1.
         






\end{thebibliography}
\end{document}